# Statistical Temperature Coefficient Distribution in Analog RRAM Array: Impact on Neuromorphic System and Mitigation Method

## AFFILIATIONS

Heng Xu[1], Yue Sun[2], Yangyang Zhu[2], Xiaohu Wang[2,a], Guoxuan Qin[1,a]

[1]School of Microelectronics, Tianjin University, Tianjin, 300072, PR China

[2]School of Mechanical Engineering, Dalian University of Technology, Dalian, Liaoning, 116024, PR China

[a] Authors to whom correspondence should be addressed: wangxiaohu@dlut.edu.cn and gqin@tju.edu.cn

## ABSTRACT

Emerging analog resistive random access memory (RRAM) based on HfOx is an attractive device for non-von Neumann neuromorphic computing systems. The differences in temperature dependent conductance drift among cells hamper computing accuracy, characterized by the statistical distribution of temperature coefficient(Tα). A compact model was presented in order to investigate the statistical distribution of Tα under different resistance states. Based on this model, the physical mechanism of thermal instability of cells with a positive Tα was elucidated. Furthermore, this model can also effectively evaluate the impact of conductance distribution of different levels under various temperatures in artificial neural networks (ANN). An approach incorporating the optimized conductance range selection and the current compensation scheme was proposed to reduce the impacts of the distribution of Tα. The simulation results showed that recognition accuracy was improved from 79.8% to 89.6% for the application of MNIST handwriting digits classification with a two-layer perceptron at 400K after adopting the proposed optimization method.

In recent years, brain inspired neuromorphic computing has demonstrated promising characteristics in terms of computing efficiency and energy consumption compared with conventional Von-Neumann architecture[1]. Non-volatile memory represented by resistive random access memory (RRAM) has been extensively studied as synaptic elements in brain-inspired computing, which is mainly used to build a high-speed and low-power neuromorphic computing system[2–7]. $HfO_x$-based RRAM recently has been widely utilized in neuromorphic hardware systems due to its fast switching speed, low power consumption, high reliability, excellent analog switching properties and great compatibility with the mainstream CMOS fabrication process[8,9].

However, the calculation accuracy of the neuromorphic computing system based on the memristor is restricted by the non-ideal effect of the memristor, such as endurance and retention degradation, read/write noises, the intrinsic nonlinearity of conductance update [10]. Therefore, the impact of device-level and array-level non-ideal effects on the accuracy of neuromorphic computing systems has been widely studied[11–14]. Meanwhile, the thermal stability of analog or multi-stage RRAM is crucial for neuromorphic terminal devices that work in a wide temperature range. Different from the well-studied binary RRAM used for storage, the resistance of the analog memristor acting as synaptic weights in neuromorphic computing system will overlap significantly with temperature changes and thus which will cause a decrease in inference accuracy[15].

Low-temperature characteristics and the impact of temperature on the reset operation of HfOx-based RRAM have been studied[16,17]. Meanwhile, the impact of operating temperature on the read/write reliability has been examined[18]. The temperature dependent statistical model has been proposed to predict the transport dependence in the temperature range below 300 K[19]. The temperature coefficient($T_\alpha$) of resistance is one of the significant indicators for evaluating thermal stability. Opposite to the high resistance state(HRS), the conduction mechanism of HfOx-based RRAM in the low resistance state(LRS) is considered to be metallic with a positive $T_\alpha$. The study found that different cells with the same resistance range in the RRAM array will show either positive or negative $T_\alpha$[20]. Therefore, the synaptic weight in the neuromorphic system will further deviate from the initial value with the temperature changes.

In this work, a compact model is proposed to predict the statistical conductance evolution with temperature changes, which can explain the physical origin of the unstable properties of cells with positive $T_\alpha$. The compact model demonstrates the excellent consistency between experimental data and simulation. Meanwhile, the impact of the statistical distribution of the $T_\alpha$ at the array level on the accuracy of neuromorphic computing systems is effectively evaluated. By selecting the conductance mapping range and the current compensation scheme, the calculation accuracy of the system can be effectively improved from 79.8% to 89.6% at 400K. This method is invalid when the temperature is lower than 350K. Therefore, maintaining excellent heat dissipation should be given priority to improve the calculation accuracy for neuromorphic systems with a temperature below 350K.

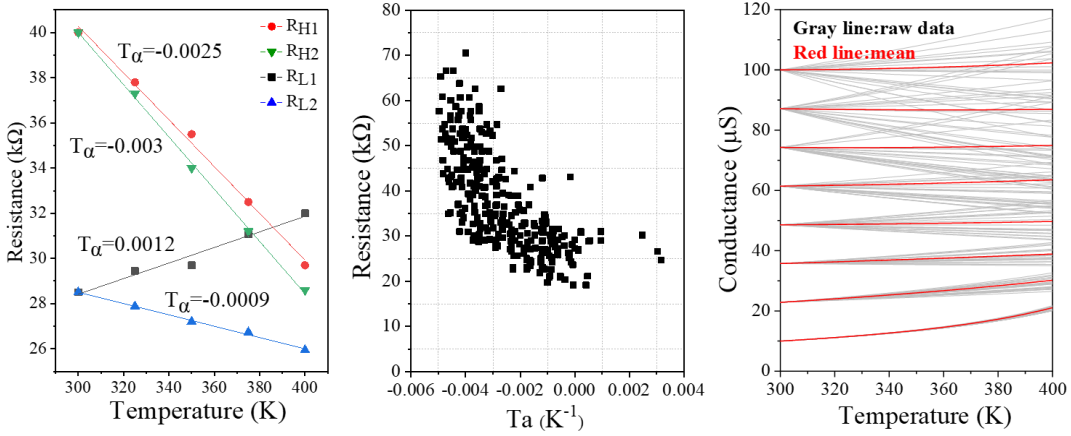

Fig. 1 (a) The temperature dependence of cell resistance with positive and negative $T\alpha$. (b) Distribution of $T_\alpha$ versus cell resistance for 1kbit array. (c) The conductance drift of 160 RRAM cells with respect to 8 conductance states as a function of temperature. Every line represents one RRAM cell.

Fig. 1(a) shows temperature dependent measurement results of two different resistance value patterns of $HfO_x$-based RRAM cells. The resistance of $R_{H1}$ and $R_{H2}$ decreasing with increasing temperature exhibited an explicit doped semiconducting behavior. However, $R_L$ indicated metallic characteristic or semiconducting behavior. The relationship between resistance and temperature can be described by the approximation below[21],

$$R(T) = R_0[1 + T_\alpha(T - T_0)] \qquad (1)$$

where $R_0$ is resistance at a reference temperature $T_0$, and $T_\alpha$ is the temperature coefficient. The statistical distribution results of $T_\alpha$ versus resistance for 1kbit $HfO_x$-based RRAM array were presented in Fig. 1(b). The conduction mechanism of a minority of low-resistance cells transformed to be metallic. The conductance in RRAM array is simply proportional to the represented weight in neural networks[22]. Accordingly, the conductance drift was examined intuitively by programming the RRAM cells randomly to eight different conductance, as shown in Fig. 1(c). It is observed that the conductance distribution was very tight at 300K and became wider as temperature increases. The overlap among neighboring conductance occurred at a lower temperature as the conductance increases, which significantly reduced the accuracy of neuromorphic calculations.

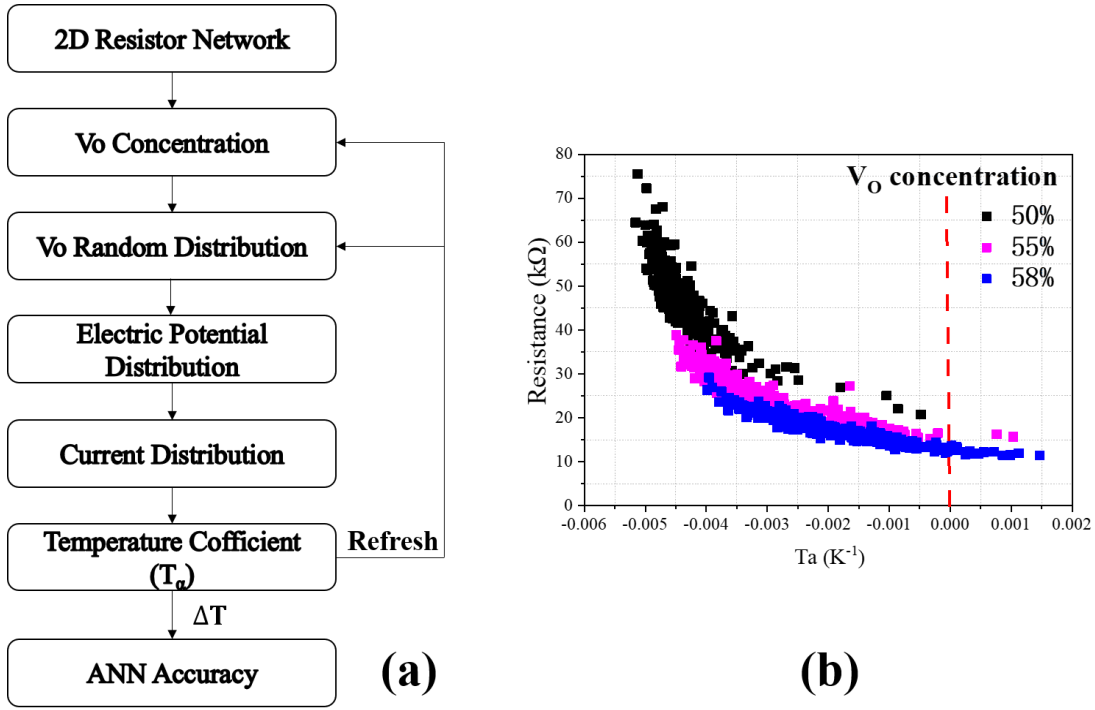

Fig. 2 (a) Simulation flow in this work. (b) Distribution of cell resistance versus $T_\alpha$ of 300 cycles of simulations with oxygen vacancy concentration of 50%, 55%, 58%, respectively.

Different from the conventional strong-filament based RRAM, in multiple-weak-filaments based RRAM, the oxygen vacancies were distributed at a nanoscale region in multiple-weak-filaments based RRAM[23]. Therefore, an atomistic simulation method was used to establish a 2D resistance network to simulate the concentration and distribution of oxygen vacancy ($V_O$) in the filament region. Fig. 2(a) shows the simulation flow. The filament region was equivalent to a stochastic distribution of different concentrations of $V_O$ in a 40×32 matrix, which was proportional to the physical dimensions of an actual device. Kirchhoff's law could solve the electric potential and current distribution in the filament area, and the overall resistance of the network can be further calculated. Similar to Conductive Bridge RAM(CBRAM), filamentary RRAM exhibits negative and positive $T_\alpha$ corresponding to conduction behavior of semiconductors and metals in high and low resistance states, respectively[24]. Therefore, different adjacent atom connection types in the 2D resistance network where $V_O$-$V_O$, $V_O$-$O^{2-}$ and $O^{2-}$-$O^{2-}$ were regarded as metallic, semiconductor and insulator resistance, respectively. Meanwhile, distinct adjacent atom connection types have corresponding temperature coefficients according to their different conductivity types.

$T_α$ was calculated using Eq. (1) to fit the overall resistance of the 2D network at different temperatures. Since the distribution of $T_α$ should be correlated with statistical results, the $V_O$ were randomly distributed at a fixed $V_O$ concentration. The above experimental process was repeated 300 times to simulate the differences of cells in the array.

The simulation result of the statistical distribution of the $T_α$ is presented in Fig. 2(b). As the oxygen vacancy concentration increases from 50% to 58%, the $T_α$ of all cells moved to a positive value, and more low-resistance cells obtained positive $T_α$. A cell with a larger resistance at the same concentration corresponds to a smaller $T_α$. The simulation result is similar to the experiment data. Under the same resistance state, the lower $V_O$ concentration value in the filament region corresponds to a higher $T_α$. For the same resistance state cells, the lower $V_O$ concentration in the filament region has a larger $T_α$. It can thus be suggested that the $V_O$ distribution in low-concentration $V_O$ cells is more concentrated and orderly to form a wider conductive path when the resistance of low-concentration cells is the same as that of the high-concentration cells.

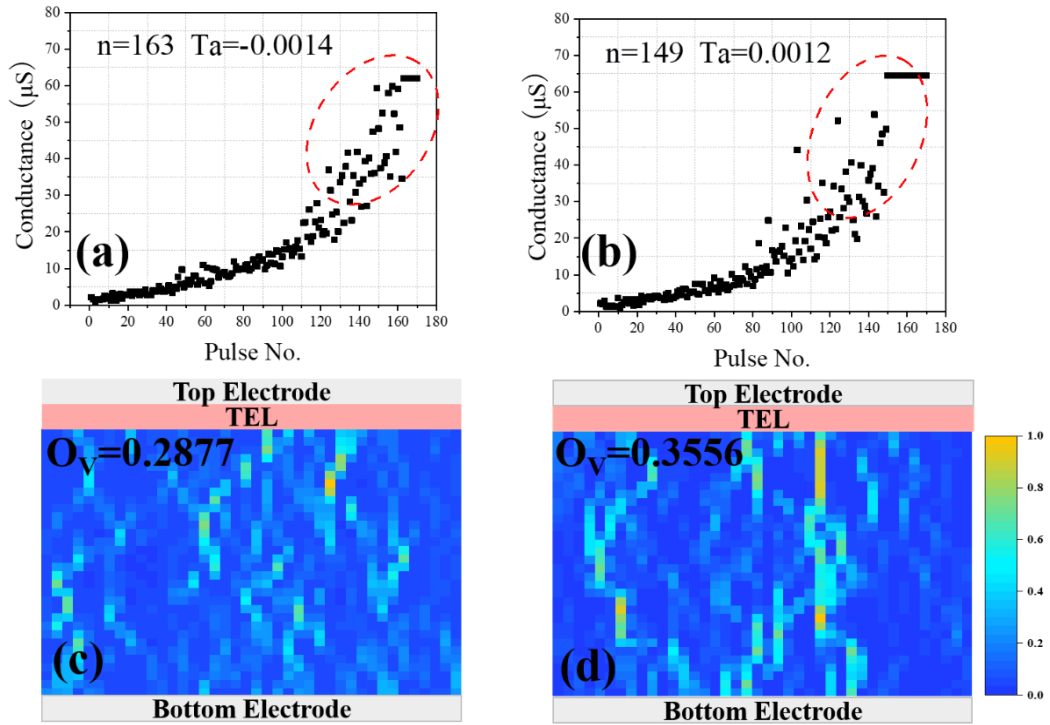

Fig. 3  Simulation results of a SET process of RRAM cells with negative (a) and positive $T_α$ (b) when voltage pulses were applied. n: the number of SET pulses. $T_α$: The temperature coefficient of the cell when the pulse stopped. Simulated distribution of the normalized current density in the filament region of $Hfo_x$ RRAM with (c) negative $T_α$ and (d) positive $T_α$.

The analog switch capability of RRAM is essential to realize the high-density weight storage of neuromorphic computing. Hence, simulate the SET process of the RRAM cells using Incremental Step Pulse Program (ISPP) scheme. The cell conductance gradually increased with the application of the SET pulse, which stopped after the device conductance exceeded the target value of 60μS. Each SET voltage pulse would increase the concentration of $V_O$ in the filament area and redistributes the $V_O$ [25]. As shown in Fig. 3(a) and Fig. 3(b), for target conductance value, the SET pulses number of negative $T_\alpha$ cell ($T_\alpha$=-0.0014) is 163, which is significantly larger than that (149) for the positive t cell ($T_\alpha$=0.0012). Meanwhile, compared with a negative $T_\alpha$ cell, the conductance of a positive $T_\alpha$ cell generally had more fluctuations after 120 pulses. Therefore, the analog switching performance of the negative $T_\alpha$ cell is preferable to that of the positive $T_\alpha$ cell. Fig. 3(c) and Fig. 3(d) show the simulated current density distribution of the negative $T_\alpha$ cell and the positive $T_\alpha$ cell. It can be seen that multiple weak CFs are formed due to the percolation effect in the negative $T_\alpha$ cell. In contrast, the apparent conductive path is formed in the positive $T_\alpha$ cell, similar to strong-filament based RRAM. An order parameter of $V_O$ can be used to evaluate the disorder effect of $V_O$ distribution[23], which can be described as

$$O_V = 2N_{V-V}/zC_V N \qquad (2)$$

where $N_{V-V}$ is the number of $V_O$-$V_O$ bond, $C_V$ is $V_O$ concentration, N is the total number of oxygen sites in the filament region, and z is coordinate number of lattice. The dispersity of $V_O$ in the filament region increases as $O_V$ decreases. $O_V$ is lower for the cell with negative $T_\alpha$ than the cell with positive $T_\alpha$.

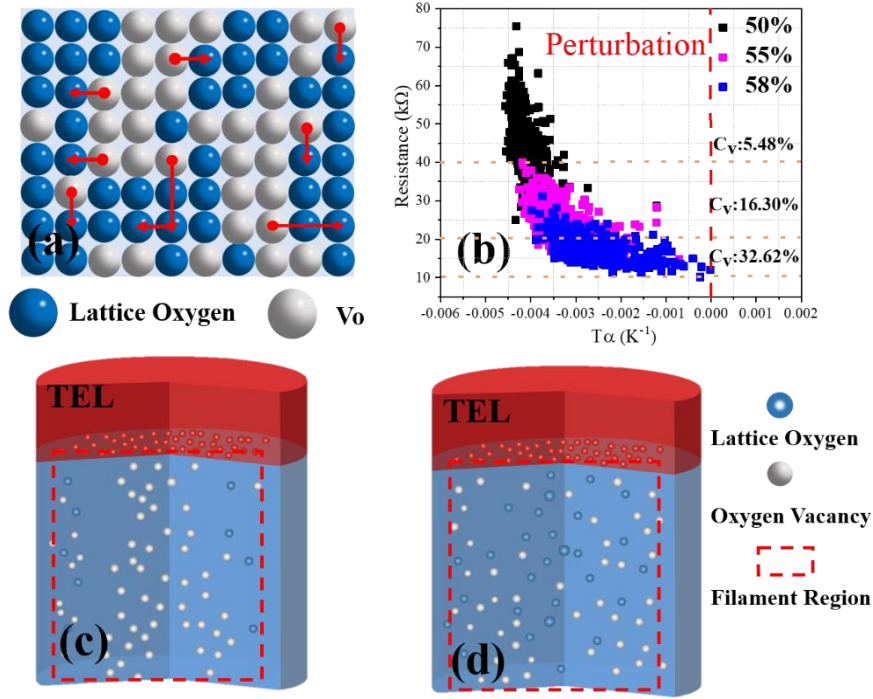

Fig. 4 (a) $V_O$ redistributes when the RRAM cell is perturbed. Most $V_O$ hop at adjacent points, but some $V_O$ can hop to farther points due to repeated reading voltage application or high temperature. (b) Redistribution of cell resistance versus $T_\alpha$. The coefficient of variation($c_v$) was used to evaluate the dispersion of the $T_\alpha$ in different resistance intervals. Schematic views of unstable positive $T_\alpha$ cell (c) and stable negative $T_\alpha$ cell (d) with thermal enhanced layer(TEL).

The retention property of cells with a positive $T_\alpha$ was much worse than that of cells with a negative value[20]. In order to further reveal the correlation between $T_\alpha$ and retention characteristics, the perturbation process of the RRAM cells was simulated. As shown in Fig. 4(a), after the cells were disturbed, the $V_O$ randomly hops at adjacent lattice sites, or $V_O$ can hop many times where this process was analogy to Brownian Motion. The simulation results of the redistribution of $T_\alpha$ are presented in Fig. 4(b). The $T_\alpha$ of cells at LRS was effectively limited to values below zero. These simulations can confirm that the strong-like filament formed in the filament region is unstable. The $V_O$ distribution in the filamentous region tended to be disordered and scattered. Therefore, repeated write-verify and heating after the programming of the RRAM can reduce the probability of cells with positive $T_\alpha$.

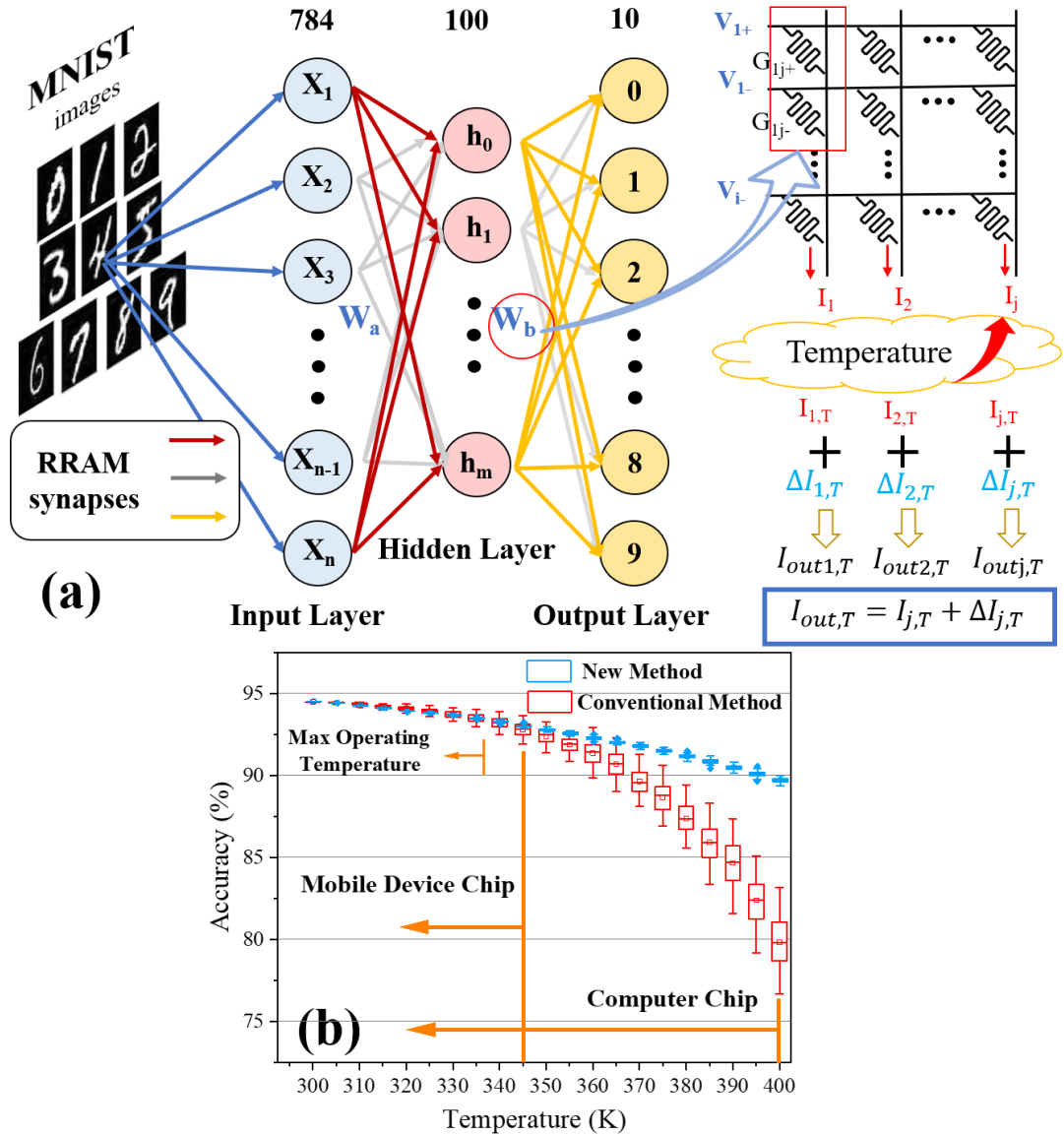

Fig. 5  (a) Schematic of a two-layer fully-connected neural network for recognizing images from the MNIST database, where the weights were implemented with the difference of two conductance in RRAM array. The final output current($I_{outj,T}$) at different temperatures can be represented by summing the output current($I_{j,T}$) of the array and the corresponding compensation current($\Delta I_{j,T}$). (b) Simulated inference accuracy of 100 RRAM chips as a function of temperature. The maximum operating temperature of the mobile device chip and the computer chip are 340k and 400k, respectively.

In order to display the potential of the compact model to evaluate and optimize neural networks, a standard multi-layer perceptron was used as an example to illustrate the influence of the distribution of $T_\alpha$ in the array on neuromorphic computing systems. This 784×100×10 fully-connected neural network was used to recognize images from Modified National Institute of Standards and Technology (MNIST) database on handwritten digits, as shown in Fig. 5(a). The activation functions of the hidden layer and the output layer were rectified linear unit (ReLU)[26] and softmax function[5], respectively. The real-valued weights were linearly mapped to the conductance difference of two RRAM cells, which was composed of the corresponding positive and negative weight rows,

$$I_j = \sum_{i=1}^{n} G_{ij} \times V_i = \sum_{i=1}^{n}(G_{ij}^+ - G_{ij}^-) \times V_i \quad (3)$$

where $G_{ij}$ is the conductance of the memory element at array position (i, j). The initial accuracy achieved using software was 97.36% (training with 32-bit single-precision floating-point weights), which was degraded to 94.48% after the quantization using 8-level weights. In order to exclusively investigate $T_\alpha$ impact on neural networks, non-ideal factors such as quantization, circuit parasitic, and retention degradation caused by temperature changes are ignored. The 8-level weight was mapped in the maximum conductance range (12.5μS-100μS). Generally, the highest operating temperature for mobile device chips is 345K, while the highest temperature for computer chips can reach 400K. When only considering the statistical distribution of $T_\alpha$, the calculation accuracy with Tα decreased from 94.48% to 79.8% for the mean when the temperature was changed from 300k to 400k.

The existing array programming methods such as write-verify and periodic-refreshing are ineffective in eliminating the impacts of $T_\alpha$ distribution on neuromorphic computing systems because the $T_\alpha$, as an intrinsic property of a material, is closely related to the device microstructure. Hence, a simple current compensation method is proposed to offset temperature-induced conductance change and recover the network performance. Firstly，a reasonable conductance mapping interval was selected according to the distribution of $T_\alpha$. The coefficient of variation ($c_v$) is used to evaluate the dispersion of the $T_\alpha$ distribution in three conductance ranges (high: 50μS-100μS, middle：25μS-50μS, and low：12.5μS-25μS),

$$c_v = \frac{\sigma}{\mu} \times 100\% \quad (4)$$

where $\sigma$ is standard deviation, and $\mu$ is mean of $T_\alpha$. The CV of $T_\alpha$ in the low conductance ranges (5.48%) was less than that in the middle (16.3%) and the high conductance ranges (32.62%). $T_\alpha$ in low conductance ranges were similar, so the cells programmed in such range can be treated as having the same $T_\alpha$(-0.004K$^{-1}$). The output current ($I_{j,T_0}$) of each column at room temperature were store in an integrated non-volatile register. The compensation current value can be calculated by measuring the operating temperature,

$$\Delta I_{j,T} = \sum_{i=1}^{n} \frac{G_i}{1 + T_\alpha \times \Delta T} \times V_i - \sum_{i=1}^{n} G_i \cdot V_i = -\frac{T_\alpha \times \Delta T}{1 + T_\alpha \times \Delta T} \times I_{j,T_0} \quad (5)$$

The recovered sum-of-product value ($I_{out,T}$) can be derived by summing actual output current ($I_{j,T}$) and compensation current ($\Delta I_{j,T}$),

$$I_{out,T} = I_{j,T} + \Delta I_{j,T} \quad (6)$$

With the optimized conductance range selection and the current compensation scheme, the accuracy distribution can be maintained at 89.6% for low-side-tail at 400K. However, it is noted that there was no difference between the inference accuracy of the compensation scheme and the initial inference accuracy with a temperature below 350K. The compensation method can reduce more accuracy loss with the temperature from 350K to 400K, but it increased the area and power consumption of the neuromorphic system. The deterioration of the cells retention behavior with increasing temperature will inevitably lead to further accuracy loss. Thus excellent heat dissipation capacity was essential for future applications of neuromorphic computing. Meanwhile, for multi-layer neural networks implemented by analog interfaces, the errors in the front layer accumulate to the next layer[27]. Compensating the output of the front layer can relieve the inference accuracy loss. More complicated neural network structures also exhibit similar characteristics.

In conclusion, thermal instability based on temperature coefficient($T_\alpha$) was comprehensively investigated in the $HfO_x$-based RRAM array. A compact model was proposed by a 2D atomistic simulation to study the statistical distribution of $T_\alpha$ on the array level. Based on the simulation of the SET process and perturbation process of the RRAM cells in the array, the physical mechanism of instability of cells with positive $T_\alpha$ was elucidated. A compensation scheme was proposed in this letter by selecting the appropriate conductance range for the weight-conductance mapping and adding compensation current, which can effectively recover the inference accuracy. With this scheme, the mean MNIST inference accuracy of a multi-layer neural network can be improved remarkably. Our results are crucial for the evaluation and optimization of RRAM-based neuromorphic computing systems.

## ACKNOWLEDGMENT

We thank Laboratory of Emerging Memory and Novel Computing of Tsinghua University for providing HfOx-based 1kbit RRAM array for testing. We acknowledge the financial support by the National Natural Science Foundation of China (Grant Nos. 61804105)and the Fundamental Research Funds for the Central Universities (Grant No. DUT19RC(3)029).

## DATA AVAILABILITY

The data that support the findings of this study are available from the corresponding authors upon reasonable request.

# REFERENCES

[1] A. Sebastian, M. Le Gallo, R. Khaddam-Aljameh, and E. Eleftheriou, Nature Nanotechnology **15**, 529 (2020).

[2] H. Lu, Y. Li, M. Chen, H. Kim, and S. Serikawa, Mobile Netw Appl **23**, 368 (2018).

[3] W. Zhang, B. Gao, J. Tang, P. Yao, S. Yu, M.-F. Chang, H.-J. Yoo, H. Qian, and H. Wu, Nat Electron **3**, 371 (2020).

[4] M.A. Zidan, J.P. Strachan, and W.D. Lu, Nature Electronics **1**, 22 (2018).

[5] F. Cai, J.M. Correll, S.H. Lee, Y. Lim, V. Bothra, Z. Zhang, M.P. Flynn, and W.D. Lu, Nature Electronics **2**, 290 (2019).

[6] C. Li, M. Hu, Y. Li, H. Jiang, N. Ge, E. Montgomery, J. Zhang, W. Song, N. Dávila, C.E. Graves, Z. Li, J.P. Strachan, P. Lin, Z. Wang, M. Barnell, Q. Wu, R.S. Williams, J.J. Yang, and Q. Xia, Nature Electronics **1**, 52 (2018).

[7] W. Zhang, **3**, 12 (2020).

[8] P. Yao, H. Wu, B. Gao, J. Tang, Q. Zhang, W. Zhang, J.J. Yang, and H. Qian, Nature **577**, 641 (2020).

[9] Q. Liu, B. Gao, P. Yao, D. Wu, J. Chen, Y. Pang, W. Zhang, Y. Liao, C.-X. Xue, W.-H. Chen, J. Tang, Y. Wang, M.-F. Chang, H. Qian, and H. Wu, in *2020 IEEE International Solid- State Circuits Conference - (ISSCC)* (IEEE, San Francisco, CA, USA, 2020), pp. 500–502.

[10] M. Zhao, B. Gao, J. Tang, H. Qian, and H. Wu, Applied Physics Reviews **7**, 011301 (2020).

[11] M. Zhao, H. Wu, B. Gao, Q. Zhang, W. Wu, S. Wang, Y. Xi, D. Wu, N. Deng, S. Yu, H.-Y. Chen, and H. Qian, in *2017 IEEE International Electron Devices Meeting (IEDM)* (IEEE, San Francisco, CA, USA, 2017), p. 39.4.1-39.4.4.

[12] P. Huang, Y.C. Xiang, Y.D. Zhao, C. Liu, B. Gao, H.Q. Wu, H. Qian, X.Y. Liu, and J.F. Kang, in *2018 IEEE International Electron Devices Meeting (IEDM)* (IEEE, San Francisco, CA, 2018), p. 40.4.1-40.4.4.

[13] P.-Y. Chen, X. Peng, and S. Yu, in *2017 IEEE International Electron Devices Meeting (IEDM)* (IEEE, San Francisco, CA, USA, 2017), p. 6.1.1-6.1.4.

[14] M. Zhao, H. Wu, B. Gao, X. Sun, Y. Liu, P. Yao, Y. Xi, X. Li, Q. Zhang, K. Wang, S. Yu, and H. Qian, in *2018 IEEE International Electron Devices Meeting (IEDM)* (IEEE, San Francisco, CA, 2018), p. 20.2.1-20.2.4.

[15] Y. Lin, C. Wang, M. Lee, D. Lee, Y. Lin, F. Lee, H. Lung, K. Wang, T. Tseng, and C. Lu, IEEE Transactions on Electron Devices **66**, 1289 (2019).

[16] R. Fang, W. Chen, L. Gao, W. Yu, and S. Yu, IEEE Electron Device Lett. **36**, 567 (2015).

[17] F.M. Puglisi, A. Qafa, and P. Pavan, IEEE Electron Device Lett. **36**, 244 (2015).

[18] T. Schultz and R. Jha, in *2019 IEEE 62nd International Midwest Symposium on Circuits and Systems (MWSCAS)* (IEEE, Dallas, TX, USA, 2019), pp. 464–467.

[19] L. Wang, A.V.-Y. Thean, and G. Liang, Appl. Phys. Lett. **112**, 253505 (2018).

[20] X.H. Wang, H. Wu, B. Gao, X. Li, N. Deng, and H. Qian, IEEE Electron Device Lett. **39**, 192 (2018).

[21] U. Russo, D. Ielmini, C. Cagli, A.L. Lacaita, S. Spiga, C. Wiemer, M. Perego, and M. Fanciulli, in *2007 IEEE International Electron Devices Meeting* (2007), pp. 775–778.

[22] T.P. Xiao, C.H. Bennett, B. Feinberg, S. Agarwal, and M.J. Marinella, Applied Physics Reviews **7**, 031301 (2020).

[23] B. Gao, H. Wu, W. Wu, X. Wang, P. Yao, Y. Xi, W. Zhang, N. Deng, P. Huang, X. Liu, J. Kang, H.-Y. Chen, S. Yu, and H. Qian, in *2017 IEEE International Electron Devices Meeting (IEDM)* (IEEE, San Francisco, CA, USA, 2017), p. 4.4.1-4.4.4.

[24] W. Guan, M. Liu, S. Long, Q. Liu, and W. Wang, Appl. Phys. Lett. **93**, 223506 (2008).

[25] W. Wu, H. Wu, B. Gao, N. Deng, S. Yu, and H. Qian, IEEE Electron Device Lett. **38**, 1019 (2017).

[26] C.-C. Chang, J.-C. Liu, Y.-L. Shen, T. Chou, P.-C. Chen, I.-T. Wang, C.-C. Su, M.-H. Wu, B. Hudec, C.-C. Chang, C.-M. Tsai, T.-S. Chang, H.-S.P. Wong, and T.-H. Hou, in *2017 IEEE International Electron Devices Meeting (IEDM)* (2017), p. 11.6.1-11.6.4.

[27] M. Hu, J.P. Strachan, Z. Li, E.M. Grafals, N. Davila, C. Graves, S. Lam, N. Ge, J.J. Yang, and R.S. Williams, in *Proceedings of the 53rd Annual Design Automation Conference* (ACM, Austin Texas, 2016), pp. 1–6.